\documentstyle[aaspp4]{article}
\def\spose#1{\hbox to 0pt{#1\hss}}
\def\simlt{\mathrel{\spose{\lower 3pt\hbox{$\mathchar"218$}}
        \raise 2.0pt\hbox{$\mathchar"13C$}}}
\def\simgt{\mathrel{\spose{\lower 3pt\hbox{$\mathchar"218$}}
     \raise 2.0pt\hbox{$\mathchar"13E$}}}
\def\ecr{\nonumber\\}

\begin{document}

\newcommand{\Phigb}{\Phi_{\gamma b}}
\newcommand{\RF}{{\cal{T}}}
\newcommand{\Aa}{{\cal A}_a}
\newcommand{\Ab}{{\cal A}_b}
\newcommand{\bg}{{b\gamma}}
\newcommand{\eal}{\!\!\! & = & \!\!\!}
\newcommand{\fract}[2]{\displaystyle{{#1 \over #2}}}

\rightline{IASSNS-AST-96/47}
 
\title{THE DAMPING TAIL OF CMB ANISOTROPIES}
\author{Wayne Hu}
\affil{Institute for Advanced Study, School of Natural Sciences, \\
Princeton, NJ 08540}
\authoremail{whu@sns.ias.edu}
\and
\author{Martin White}
\affil{Enrico Fermi Institute, University of Chicago, \\ Chicago, IL 60637}
\authoremail{white@oddjob.uchicago.edu}

\begin{abstract}
\noindent
\rightskip=0pt
By decomposing the damping tail of CMB anisotropies 
into a series of transfer functions representing individual
physical effects,
we provide ingredients that will aid in the reconstruction of the
cosmological model from small-scale CMB anisotropy data. 
We accurately calibrate the model-independent effects of diffusion and
reionization damping which provide potentially the most robust information on
the background cosmology.
Removing these effects, we uncover model-dependent processes such as the
acoustic peak modulation and gravitational enhancement that can help
distinguish between alternate models of structure formation and provide
windows into the evolution of fluctuations at various stages in 
their growth. 
\end{abstract}

\keywords{cosmology:theory -- cosmic microwave background. \\ 
{\it Submitted: }September 10, 1996 to ApJ}

%\tableofcontents
\rightskip=0pt
\section{Introduction}

Much effort is being expended to measure the angular power spectrum of the
cosmic microwave background (CMB) anisotropy on increasingly smaller 
angular
scales.  For many types of models for structure formation, 
the spectrum can be predicted
to a precision of about 1\% (\cite{HSSW}), raising the
hope that the cosmological parameters that are the input to these
calculations can be extracted to comparable precision 
(see e.g.~\cite{JunPRD}).
The ``inverse problem'' of reconstructing the model given a spectrum is 
less
well understood than the ``forward problem'' of predicting it 
given the model.
For this purpose,
it is important to assess the generation of anisotropies in
a manner that is not tied to any given model for structure formation.  
{}From the theory of anisotropy formation, we know that CMB fluctuations
suffered causal processing and damping of the primordial signal.
In this paper, we numerically calibrate such effects, 
extending and improving upon prior work
(Hu \& Sugiyama 1995ab, 1996, hereafter \cite{HSa,HSb,HSc} 
and \cite{HWLong}). 

A particularly fruitful way to visualize the CMB spectrum, and one that
provides a framework for the inverse problem, is as a product of 
transfer functions representing individual physical
effects.
The spectrum is then constructed out of physical elements 
rather than a model-dependent
parameterization. 
Conceptually, the evolution of CMB anisotropies processes 
primordial metric or
gravitational potential perturbations into features observable in the spectrum
today (see e.g.~\cite{Bond,HWLong}).  Since the evolution obeys linear
perturbation theory, 
its effects are described by a series of transfer functions
that multiply the underlying perturbations.
The form of these functions depends on the cosmological model, not 
only for the
background expansion and thermal history (see e.g.~\cite{Bonetal,Sel,HSS}) but
also for structure formation (see e.g.~\cite{CriTur,Magetal,Duretal,HSW}).
By decomposing the evolution into functions 
representing separate physical
effects, we can isolate portions of the anisotropy spectrum 
that are the most
sensitive to particular aspects of the cosmological model.  

In particular, processes that damp CMB anisotropies, photon diffusion
(\cite{Sil}) and rescattering (\cite{EfsBon}), depend mainly on the background
parameters and little on the perturbations that form structure in the universe.
In \S \ref{sec-calculation}, we isolate these effects in a numerical treatment.
{}From this damping calibration, we produce convenient fitting formulae that
accurately describe the behavior of the diffusion and reionization damping
transfer functions, or envelopes, directly in anisotropy multipole space.
In \S \ref{sec-information}, we illustrate the reconstruction 
process by testing it with
known models within the cold dark matter (CDM) scenario.  By removing the 
model-independent effects of damping, one uncovers important model-dependent
effects such as the baryon-drag modulation of the peaks (\cite{HSa}), the
potential envelope that describes gravitational driving of acoustic
oscillations (\cite{HSc}), and the regeneration of anisotropies during
reionization (\cite{SZ,Kai84}).

In the context of currently popular models, recovery of these signatures will
help distinguish between such possibilities as an inflationary or cosmological
defect origin of fluctuations (\cite{CriTur,Duretal,HWLong}).  
The effects of damping are also 
intrinsically interesting because they provide the most 
model-independent probes of the background cosmology.  
We also consider how diffusion
damping can be used to constrain the curvature of the universe and reionization
damping to determine the redshift and extent of reionization in the universe.
In this way, the study of effects in the damping tail of 
CMB anisotropies presented here
will aid in the future reconstruction of the cosmological model from the
anisotropy data.  

\section{Damping Calculation}
\label{sec-calculation}

Damping processes which affect CMB anisotropies provide the most 
model-independent information available in the spectrum and allow
constraints on cosmological parameters such as the curvature 
and the thermal history of the universe. 
Furthermore, these universal effects obscure the
model-{\it dependent\/} signatures that are useful to determine the mechanism
for structure formation in the universe and the ultimate source of density 
perturbations.  

For both these reasons, an accurate calibration of
damping effects is desirable.  
In this section, we begin with the formalism necessary to
describe them (\S \ref{ss-boltzmann}) and simple approximations to help
understand their nature (\S \ref{ss-analytic}).
We then turn to numerical calibration of these effects (\S \ref{ss-numerical}).
Finally, we give convenient fitting formulae to their effects on the
anisotropy power spectrum that encapsulate these results (\S \ref{ss-fitting}).

\subsection{Boltzmann Formalism}
\label{ss-boltzmann}

In this section, we provide the formalism for the evolution of CMB
anisotropies that underly the calculations which follow.
It may be skimmed upon first reading.

The anisotropy in the CMB is described by small perturbations of the
photon distribution function around a homogeneous and isotropic
black-body.  The Boltzmann equation describes the evolution
of the distribution function $f$, through Compton scattering
with electrons $df/d\eta\, (\eta,{\bf x}(\eta),{\bf p}(\eta)) = C[f]$,
where the collision term is written schematically as $C[f]$.
Here $\eta$ is the conformal time and ${\bf p}$ the photon momentum.
In the absence of spectral distortions, the magnitude of the
momentum can be integrated over leaving only its directional
depenence ${\bf \gamma}$ and the effect of gravitational
redshifts on the photon temperature perturbation $\Theta$.
Due to azimuthal symmetry and the decoupling of modes in
linear theory, it is convenient to decompose the fluctuation in
a Fourier or normal mode $k$ into angular moments,
e.g.~in flat space
$\Theta(\eta,{\bf k},{\bf \gamma}) = \sum_\ell (-i)^\ell \Theta_\ell
P_\ell({\bf k} \cdot {\bf \gamma})$
with an appropriate
generalization to curved spaces (\cite{Wil,WhiSco}).  
Here $\gamma_i$ are the direction cosines of the photon momenta. 
The Boltzmann equation then becomes an infinite hierarchy of 
coupled ordinary differential equations (see e.g.~\cite{MaBer})
\begin{eqnarray}
\dot \Theta_0 \eal -{k \over 3} \Theta_1 - \dot \Phi, \ecr
\dot \Theta_1 \eal k \left[\Theta_0 + \Psi - {2 \over 5}\kappa_2 \Theta_2 \right]
                 - \dot \tau (\Theta_1 - v_b), \ecr
\dot \Theta_2 \eal k \left[{2 \over 3}\kappa_2 \Theta_1 - {3 \over 7} 
\kappa_3 \Theta_3 \right]
                 - \dot \tau \left( {9 \over 10} \Theta_2 - {1 \over 10}
                Q_2 - {1 \over 2} Q_0 \right), \ecr
\dot \Theta_\ell \eal k \left[ {\ell \over 2\ell -1} \kappa_\ell 
		\Theta_{\ell - 1} - {\ell +1 \over 2\ell + 3} 
		\kappa_{\ell+1} \Theta_{\ell + 1} \right]
                      - \dot\tau \Theta_\ell,\quad {(\ell > 2)}
\label{eqn:boltzmann}
\end{eqnarray}
where $\kappa_\ell = [1-(\ell^2-1)K/k^2]^{1/2}$ modifies the angular
hierarchy for geodesic deviation in spaces of 
constant comoving curvature 
$K = -H_0^2 (1-\Omega_0-\Omega_\Lambda)$ with a Hubble
constant of $H_0 = 100 h$ km s$^{-1}$ Mpc$^{-1}$.  
The metric perturbations are represented by $\Phi$ the fluctuation
of the spatial
curvature in Newtonian gauge and $\Psi$ the Newtonian potential.
The collision terms from $C[f]$ are proportional to
$\dot \tau = n_e \sigma_T a$ the differential optical depth to
Compton scattering, where $n_e$ is the free electron density and
$\sigma_T$ is the Thomson cross section.

Scattering by electrons
with velocity $v_b$ generates a Doppler effect on the photons.
Scattering of anisotropic radiation creates a polarization,
described by the temperature perturbation in the Stokes parameter $Q$,
and governed by a separate hierarchy (\cite{BE84}),
\begin{eqnarray}
\dot Q_0 \eal -{k \over 3} Q_1 - \dot \tau
                    \left[ {1 \over 2} Q_0 - {1 \over 10}
                    (\Theta_2 + Q_2) \right], \ecr
\dot Q_1 \eal k \left[Q_0 - {2 \over 5} \kappa_2 Q_2 \right]
                 - \dot \tau \Theta_1, \ecr
\dot Q_2 \eal k \left[{2 \over 3} \kappa_2 Q_1
                 - {3 \over 7} \kappa_3 Q_3 \right]
                 - \dot \tau \left( {9 \over 10} Q_2 - {1 \over 10}
                \Theta_2 - {1 \over 2} Q_0 \right), \ecr
\dot Q_\ell \eal k \left[ {\ell \over 2\ell -1} \kappa_\ell Q_{\ell - 1}
                      - {\ell +1 \over 2\ell + 3} \kappa_{\ell+1}
			Q_{\ell + 1} \right]
                      - \dot\tau Q_\ell, \quad {(\ell > 2)}.
\label{eqn:polarization}
\end{eqnarray}

To complete these equations, we need the baryon Euler equation, 
which determines the evolution of the baryon velocity, 
\begin{equation}
\dot v_b = -{\dot a \over a} v_b + k\Psi + \dot\tau(\Theta_1-v_b)/R.
\label{eqn:baryoneuler}
\end{equation}
%and the Einstein-Poisson equations to relate the metric perturbations
%to the energy-momentum tensor or density, velocity, 
%isotropic and anisotropic stress  perturbations of the
%matter, $\delta,$ $v$,
%$\delta p/p$, $\pi$ respectively, to the metric perturbations
%\begin{eqnarray}
%(k^2 - 3K)\Phi \eal  4\pi G a^2 \sum [ \rho_i \delta_i + 3(\rho_i + p_i)
%{\dot a \over a} v_i/k], \ecr
%k^2 (\Psi + \Phi) \eal -8\pi Ga^2 \sum p_i \pi_i,
%\label{eqn:poisson}
%\end{eqnarray}
%where the sum is over particle species and $\pi$'s are the anisotropic
%stress perturbations. 
Finally, the observable anisotropy spectrum
follows by integrating over the $k$-modes,
\begin{equation}
	{{2\ell + 1} \over 4\pi}C_\ell =
	{1 \over 2 \pi^2} \int {dk \over k} {k^3 |\Theta_\ell(\eta_0,k)|^2
	\over 2\ell +1 }.
\end{equation}

The interpretation of these equations is quite straightforward.  The metric
fluctuations feed power into hierarchy through the gravitational 
redshift effects of density dilution ($\dot\Phi$ in $\ell=0$)
and potential infall ($k\Psi$ in $\ell=1$). 
If the optical depth across a wavelength $\dot{\tau}/k\ll1$,
this power flows to higher $\ell$ 
much like a wave pulse
flows along a string, being concentrated in mode $\ell$ 
when $k\eta\sim\ell$. The critical epoch for this process is 
horizon crossing $k\eta \sim 1$ after which $\ell \simgt 1$ 
modes can be populated.
When the free electron density is not negligible then 
the Compton scattering
terms ($\dot{\tau}$ terms) become important.
Modes with $\ell\ge2$ are exponentially damped, sealing
off the hierarchy and providing a barrier
off which the wave pulse reflects.  The monopole term
is not damped at all and the dipole term is driven
toward $v_b$ so that the distribution is isotropic in the electron rest frame.

Thus before recombination, $\dot{\tau}/k\gg1$, and the photon distribution
possesses only the $\ell=0$ (density) and $\ell=1$ (velocity) modes, which
represent a fluid that oscillates acoustically due to photon pressure
(see \S\ref{ss-analytic}).
Only for very high $k$ will power leak into the higher $\ell$ modes, where it
will be exponentially damped.
This is responsible for the damping tail at small angular scales.
An increase in $\dot{\tau}$ at late times due to reionization also possesses
a characteristic signature.
For scales inside the horizon at reionization, the power has already
propagated to high $\ell$ where it suffers exponential damping;
for larger scales no such damping occurs.
Thus reionization damps small-scale anisotropies while
preserving large-scale anisotropies.
We shall discuss these behaviors more quantitatively in the next section.

\subsection{Analytic Estimates}
\label{ss-analytic}

Before turning to the numerical calibration of effects in the
damping tail, it is useful to describe them analytically to see
how they enter in and affect the Boltzmann evolution given above.
The two main damping processes at work in the CMB are photon diffusion
before recombination and rescattering during an epoch of late 
reionization.

\subsubsection{Diffusion Damping before Recombination}

For wavelengths much larger than the mean free path to Compton 
scattering $(k/\dot\tau \ll 1)$,
the Boltzmann hierarchy of Eq.~(\ref{eqn:boltzmann}) can be described by the
relativistic fluid dynamics of a combined photon-baryon fluid.
Rapid scattering insures that any anisotropy of the photons in the electron
rest frame is vanishingly small so that the hierarchy can be truncated at
$\ell=1$ with $\Theta_1 \approx v_b$.
Even in this {\it tightly-coupled\/} regime, the random walk of the photons
through the electrons eventually mixes photons across a wavelength of the
fluctuation (\cite{Sil}). 
Thus, we expect temperature perturbations to be destroyed by this diffusive
process {\it before} the mean free path grows long enough to invalidate the
central approximation.
This statement is only approximately true during the recombination epoch when
the mean free path grows so rapidly that it approaches the horizon scale and
coincides with the diffusion scale.  For this reason, we calibrate the
process numerically in \S\ref{ss-numerical}.

Formally, we can approximate these effects by expanding the Boltzmann
temperature and polarization equations (\ref{eqn:boltzmann}) and 
(\ref{eqn:polarization}) in powers of $k/\dot\tau$ (\cite{PeeYu}).  
To lowest order, one obtains the oscillator equation (\cite{HSa})
\begin{equation}
{d \over d\eta}(1+R)\dot\Theta_0 + {k^2 \over 3}\Theta_0 =
- {k^2 \over 3}(1+R)\Psi - {d \over d\eta}(1+R)\dot\Phi.
\label{eqn:oscillator}
\end{equation}
Gravity drives the oscillator by potential infall into $\Psi$ and
density dilution as the curvature fluctuation  $\Phi$ changes.
The baryon inertia in the fluid is described by the relative baryon-photon
momentum density ratio $R$ and increases the effective mass of the oscillator.
Together these effects imply oscillations at the sound speed
\begin{equation}
c_s = {1 \over \sqrt{3(1+R)}},
\label{eqn:sound}
\end{equation}
around a zero point displaced by gravity to $(1+R)\Psi$ for slowly-varying
$\Phi$ (\cite{HSa,HSc,HWLong}).

To treat the effects of diffusion, one must include higher order terms.
An examination of the $\ell=1$ photon Euler equation (\ref{eqn:boltzmann})
shows that there are two diffusive effects: viscous damping from the
quadrupole $\Theta_2$ and heat conduction from the relative photon-baryon
velocity $\Theta_1 - v_b$ (\cite{Wei}).  
From the expansion of the polarization hierarchy,
$Q_2 = Q_0 = {1 \over 4}\Theta_2$ and the quadrupole evolution equation 
(\ref{eqn:boltzmann}) with $\Theta_1 \gg \Theta_3$, we obtain the
tight-coupling prediction for the quadrupole,
\begin{equation}
\Theta_2 = (k/\dot\tau){8 \over 9} \Theta_1.
\label{eqn:quadrupole}
\end{equation}
Heat conduction may be described by expanding the baryon Euler
equation (\ref{eqn:baryoneuler}) to second order.
Let us assume a solution of the form $\Theta_1\propto\exp i\int\omega d\eta$
and ignore variations on the expansion time scale $\dot{a}/a$ in comparison
with those at the oscillation frequency $\omega$.  
We return to consider this approximation in \S\ref{ss-assumptions}. 
The heat conduction equation becomes
\begin{equation}
\Theta_1 - v_b = \dot \tau^{-1} R [i\omega \Theta_1 - k\Psi] + 
	\omega^2 \dot\tau^{-2} R^2 \Theta_1,
\label{eqn:slippage}
\end{equation}
allowing us to rewrite the photon Euler equation (\ref{eqn:boltzmann}) as
\begin{equation}
i\omega (1 + R)\Theta_1 = k[\Theta_0 + (1+R)\Psi] - 
	\omega^2 \dot\tau^{-1} R^2 \Theta_1 - 
	{16 \over 45} k^2 \dot\tau^{-1} \Theta_1.
\label{eqn:eulerphoton}
\end{equation}
The presence of $(1+R)\Psi$ again reflects the gravitational zero-point
displacement of the oscillator.  It is thus appropriate to try a solution
of the form $\Theta_0 + (1+R)\Psi \propto \exp i\int \omega d\eta$.
With this assumption, one obtains the dispersion relation for acoustic
oscillations  
\begin{equation}
\omega = \pm k c_s  + i{1 \over 6} k^2 \dot \tau^{-1} 
 \left[ {R^2 \over (1+R)^2} + {16 \over 15} {1 \over 1+R}
 \right].
\label{eqn:dispersion}
\end{equation}
From the form of the solution $\exp i\int \omega d\eta$, this gives the
damping scale $k_D$ (\cite{Kai})
\begin{equation}
k_D^{-2} = \fract{1}{6} \int d\eta \, \fract{1}{\dot \tau}
	\fract{R^2 + 16(1+R)/15}{(1+R)^2}
\label{eqn:diffusionlength}
\end{equation}
by which acoustic oscillations are damped exponentially
as $\exp[-(k/k_D)^2]$. 
Notice that the diffusion length is roughly the geometric mean of the mean
free path $\dot\tau^{-1}$ and horizon length $\eta$ as one would expect
of a random walk $k_D \sim \sqrt{\dot\tau/\eta}$.

\subsubsection{Diffusion Damping during Recombination}

As the universe recombines, the mean free path and hence the diffusion length
of the photons increases.  As long as the diffusion length is much greater than
the mean free path, damping can be described by the tight-coupling
approximation of the previous section.  This is because the mean free path only
surpasses the wavelength {\it after\/} diffusion has already destroyed the
perturbations, resulting in no contributions outside of the tight-coupling
regime. 
The approximation thus remains approximately true until quite near the end of
recombination when the mean free path becomes comparable to the horizon and so
the diffusion length (\cite{HSa,HSc}).  This fact explains the reasonable level
of agreement between the numerical results we present in \S\ref{ss-numerical}
and the tight-coupling approximation.

The remaining subtlety is that due to the finite duration of recombination,
last scattering takes place at a slightly different epoch, with a slightly
different diffusion length, for each photon.
The net effect has been approximated (\cite{HSa}) by weighting the damping by
the visibility function $\dot\tau e^{-\tau}$, the probability of last
scattering within $d\eta$ of $\eta$,
\begin{equation}
{\cal D}(k) = \int_0^{\eta_0} d\eta \  (\dot \tau e^{-\tau})
	\exp\{-{[k/k_D(\eta)]}^2 \}.
\label{eqn:diffusionenva}
\end{equation}
This ``smearing'' of the last scattering surface, and the evolution of $k_D$
tend to soften the damping, meaning it is not quite the simple exponential
one would naively predict.  It is however often convenient to 
define the last scattering epoch as $\tau(\eta_*)=1$.

Notice that the net result depends only on the cosmological parameters
of the background.
The effect of $\Omega_0 h^2$ is simple.  Increasing $\Omega_0h^2$ decreases the
horizon at last scattering thus monotonically decreasing the diffusion length.
The dependence on $\Omega_bh^2$ is more complicated.
Increasing $\Omega_b h^2$
\begin{enumerate}
\item decreases the mean free path
\item delays recombination
\item shortens its duration
\item speeds diffusion scale growth at recombination
\end{enumerate}
The first effect tends to decrease the damping length and dominates
for low $\Omega_b h^2$.  The second effect extends the amount of time
the photons can diffuse and hence increases the damping length; it dominates
at high $\Omega_b h^2$.
In the limit of instantaneous recombination, the damping function ${\cal D}(k)$
attains its sharpest form of $\exp\{-{[k/k_D(\eta_*)]}^2\}$.
For the realistic case where recombination takes place over an extended period
${\cal D}(k)$ becomes less steep.
Both the width of the visibility function and the evolution of $k_D$ through it
affect this drop.  Again the baryon dependence of these two effects are in
opposition leading to a steepening of the slope at both the high and low
$\Omega_b h^2$ limits.

\subsubsection{Free Streaming}

After recombination photons enter the free-streaming regime.
The observer views a temperature fluctuation at wavenumber
$k$ on the last scattering surface as an anisotropy at multipole moment 
$\ell\approx f kr_\theta(\eta_*)$, where the constant of proportionality
$f\approx 0.98$ from numerical fitting (see \S \ref{ss-numerical}) and the
comoving angular-size distance to the epoch $\eta$ is
\begin{equation}
r_\theta(\eta) = |K|^{-1/2} \sinh[|K|^{1/2}(\eta_0-\eta)],
\label{eqn:angularsize}
\end{equation}
for $K<0$.  For positively curved universes, replace sinh with sin.
The fact that angular size depends sensitively on the curvature allows
its precise measurement from acoustic features
(\cite{DZS,SugGou,KamSS,HWLong}).
These effects are 
represented formally in the Boltzmann equation (\ref{eqn:boltzmann})
by the transfer of power down the $\ell$-hierarchy with distance 
$\eta-\eta_*$
from the last scattering event and the geodesic deviation factors
$\kappa_\ell$.
Note that the latter becomes important when the distance is long 
enough such that the subtended angle $\theta \sim \ell^{-1} =
{\sqrt{K}/k}$, i.e.~smaller than that of a wavelength at the
curvature distance.  

The angular-size distance relation may be used to map 
$k$-space inhomogeneities
onto $\ell$-space anisotropies.
For example, the damping function in multipole space is
\begin{equation}
{\cal D}_\ell \approx {\cal D}(k=\ell/ f r_\theta(\eta_*)).
\label{eqn:projection}
\end{equation}
There are instances when this mapping fails to accurately describe the
streaming process.  The projection of $k$-mode inhomogeneities onto
$\ell$-mode 
anisotropies depends on the viewing angle and is thus not one-to-one.
In particular, it can take power to larger angles for wavelengths that happen
to be viewed with wavevector parallel to the line of sight.  
In this case, the angular separation between the intersections of the flat
wavefront with the spherical shell at $\eta_*$ is much larger than
$(kr_\theta)^{-1}$. 
Sharp features in $k$-space will thus be blurred in $\ell$-space and excess
power at small physical scales can be aliased into large angular scales.
Formally, this is reflected by the decomposition of the $k$-mode on the sphere
and the fact that the solution to the sourceless Boltzmann or Liouville 
equation is just its radial component, a spherical Bessel function in flat
space (see e.g.~\cite{BE87}).  
For the cases we consider, where the $k$-space features are broad with no
strong deviations from scale invariance, the simple approximation of
Eq.~(\ref{eqn:angularsize}) suffices.

\subsubsection{Reionization Damping}

From the null detection of the Gunn-Peterson effect (\cite{GunPet}), 
we know that the universe was reionized at least as early as redshift 
$z\approx 5$.  
Reionization recouples the photons to the electron-baryon plasma.
The same process that is responsible for diffusion damping acts
to destroy anisotropies during this epoch as well.

During the free-streaming epoch, the effective ``diffusion length'' 
is simply the horizon scale.  Photon trajectories from different
temperature regions on the last scattering surface intersect
forming the anisotropy that is represented by the
$\ell \ge 2$ photon modes. 
When the universe reionizes, the photons which rescatter lose
their anisotropy.  Note that the isotropic temperature fluctuation
that exists above the horizon, where trajectories have not yet
crossed, does not damp by rescattering.  
This is reflected in the lack of a $\dot\tau$ coupling term in
the $\ell=0$ mode of the Boltzmann equation (\ref{eqn:boltzmann}).  
The $\ell=1$ mode damps in such a way as to drive $\Theta_1$ toward
$v_b$ so that the distribution is isotropic in the electron rest frame. 

Even in a reionized universe, photons eventually
last scatter as the electron density drops due to the expansion and
the mean free path to scattering exceeds the horizon length.  
Thus only the fraction $e^{-\tau}$ of the photons that did not
rescatter contribute to the anisotropy below the horizon at last
scattering $\eta_r$.  Above this scale all photons contribute. 
Thus the rough form of the reionization damping 
function becomes
\begin{equation}
{\cal R}_\ell = \cases{ 1 & $\ell \ll r_\theta /\eta_r$, \cr
			e^{-\tau} & $\ell \gg r_\theta/ \eta_r$, \cr }
\label{eqn:reionenva}
\end{equation}
where again the effect of the finite duration of last
scattering on $\eta_r$ can be accounted for by the
visibility function (see \S \ref{ss-fitting}).

\begin{figure}[t]
\begin{center}
\leavevmode
\epsfxsize=3.5in \epsfbox{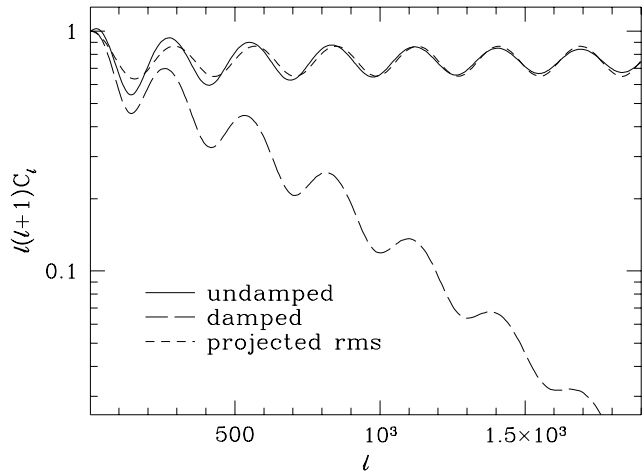}
\end{center}
\caption{Diffusion damping calibration.  In the absence of both 
diffusion damping and gravitational sources, the rms temperature
fluctuation at recombination (short-dashed line)
exhibits simple acoustic oscillations.
These are mapped onto anisotropies on the sky in a near one to
one fashion (solid line).  The inclusion of diffusion terms in the
Boltzmann equation allows for a simple numerical calibration of its 
effects.}
\label{fig:damp}
\end{figure}

\subsection{Numerical Calibration}
\label{ss-numerical}

The expressions of the previous section are only approximations, though
useful ones.  We now turn to numerical calibration by solving the
Boltzmann equations of \S\ref{ss-boltzmann}.  

Extracting the damping effects from realistic models of structure formation is
complicated due to the manner in which gravity generates perturbations through
the metric fluctuations $\Phi$ and $\Psi$ in the model. 
Since the effects discussed above are essentially model-independent, we choose
instead to calculate a toy-model in which no gravitational effects, 
beyond the background expansion, are included.  
Specifically, we solve the Boltzmann equations for the photons and baryons
with $\Phi=0=\Psi$. 
This includes neglecting the self-gravity of the photon-baryon fluid.  

Before recombination, we are left with a pure acoustic oscillation 
whose behavior is completely determined by the initial conditions.  
For simplicity, we take them to be adiabatic and scale invariant. 
The evolution equations of \S \ref{ss-boltzmann} 
are then solved in the usual way (see e.g.~\cite{BE84,MaBer,SelZal})
through recombination to the present.
This properly includes the effects of diffusion through the last 
scattering surface and the
projection of the fluctuations at last scattering onto the sky today.
We show an example in Fig.~\ref{fig:damp} (long-dashed line).

To extract the diffusion damping behavior, we compare this to a calculation
of the same model with diffusion damping ``turned off''.
Specifically, we solve the tightly-coupled photon-baryon equation
(\ref{eqn:oscillator}) up to the point where the optical depth to the present,
(ignoring reionization) becomes unity.
We then free-stream the photons to the present by solution of 
the sourceless
Boltzmann equation ($\Psi=\Phi=\dot\tau=0$ in Eqs.~\ref{eqn:boltzmann}
and \ref{eqn:polarization}) to determine the anisotropy
(see Fig.~\ref{fig:damp}, solid line).
The ratio of the angular power spectrum of the damped to undamped calculation
gives the form of the damping function ${\cal D}_\ell^2$
(see also Fig.~\ref{fig:scdm}).
This also serves to calibrate the angular-size distance relation of
Eq.~(\ref{eqn:angularsize}) through comparison with the mean squared
fluctuations in Fourier space $|\Theta_0|^2+|\Theta_1|^2/3$ at optical
depth unity (see Fig.~\ref{fig:damp}, short-dashed line).
By aligning the peaks, one extracts the proportionality factor $f\approx0.98$.
As discussed in \S \ref{ss-analytic} free streaming smears features in the
$k$-space rms spectrum somewhat which explains the slightly smoother actual
anisotropy. 

\begin{figure}[t]
\begin{center}
\leavevmode
\epsfxsize=3.5in \epsfbox{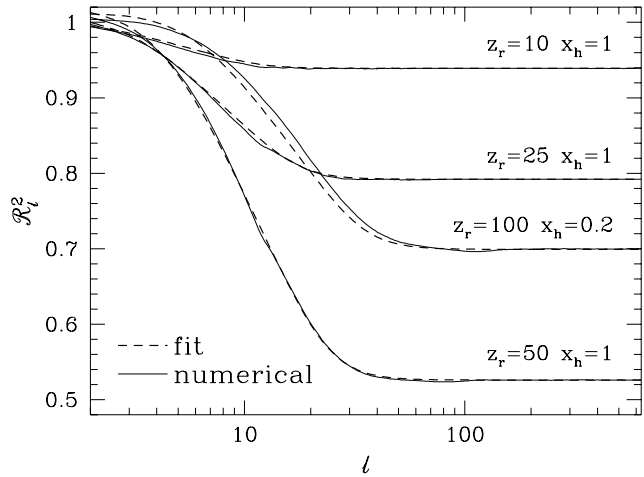}
\end{center}
\caption{Reionization damping calibration. By removing the relative
Doppler effect from a reionized Boltzmann calculation and comparing
the result to the same model (here
standard CDM $\Omega_0=1$, $h=0.5$, $\Omega_b h^2=0.0125$) with no
reionization, the effects of rescattering damping are isolated.
The reionization damping envelope is fit by two parameters,
the optical depth during reionization and the horizon scale at
last scattering (see Eq.~24).
%CAUTION FIXED NUMBER
}
\label{fig:reion}
\end{figure}

To extract the reionization damping behavior, we compare the no-reionization
case to one with some arbitrary reionization history.  In order to 
isolate damping effects from the Doppler effect due to the relative
motion of the baryons with respect to the CMB, we set $v_b=\Theta_1$ 
during this epoch.  For simplicity, we often parameterize the reionization
as instantaneous at some epoch $z_r$ to some constant fractional level of
hydrogen reionization $x_h$ though none of our results depend on this
simplification.  The ratio of the two power spectra gives 
${\cal R}_\ell^2$.  We show examples in Fig. \ref{fig:reion} (solid lines).

\subsection{Fitting Formulae}
\label{ss-fitting}

It is convenient to fit the numerical calculations of \S\ref{ss-numerical} for
the diffusion damping and reionization damping envelopes. Aside from providing 
a compact summary of the results, this exposes the sensitivity of the spectrum
to cosmological parameters which will be useful in \S\ref{sec-information}.

\subsubsection{Diffusion Damping Envelope}
 
Since the effect of diffusion damping is determined solely through the Compton
mean free path and horizon scale, it is dependent on very few cosmological
parameters.
The Compton mean free path of a photon is governed by the baryon density
$\rho_b \propto \Omega_b h^2$.  If the present energy density in the radiation
is fixed, then the horizon only depends on the matter content $\Omega_0 h^2$
before curvature and cosmological constant contributions become significant.
We assume here that the radiation energy density is fixed by the observed
CMB temperature $T_\gamma = 2.728$K (\cite{Fixetal}) and there exist three
families of massless neutrinos with $T_\nu = (4/11)^{1/3}T_\gamma$ 
(we ignore the small correction of \cite{DodTur}).
Thus aside from the projection effects from $r_\theta$, which 
are sensitive to the curvature and cosmological constant, 
the damping behavior depends only
on $\Omega_0 h^2$ and $\Omega_b h^2$.
We have computed ${\cal D}_\ell$ as described in \S\ref{ss-numerical} for
150 models in the range
$0.02 < \Omega_0 h^2 < 0.75$ and $0.005 < \Omega_b h^2 < 0.75$ .

\begin{figure}[t]
\begin{center}
\leavevmode
\epsfxsize=3.5in \epsfbox{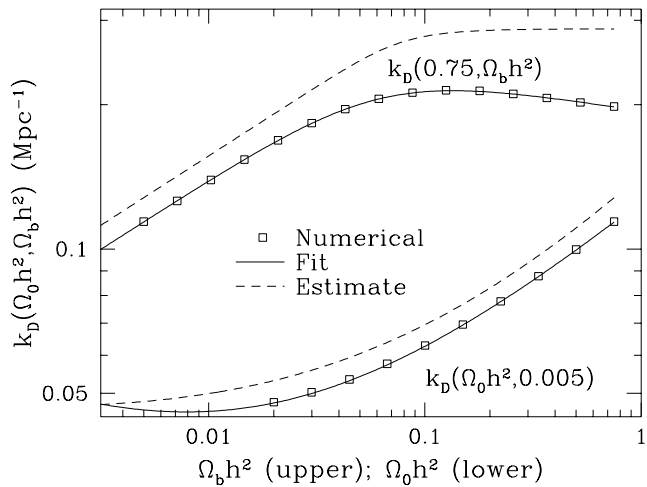}
\end{center}
\caption{Diffusion scale calibration.  Analytic estimates of
$k_D(\Omega_0 h^2, \Omega_b h^2)$
based on the tight-coupling approximation trace the results to
reasonable accuracy and explains their general behavior 
(HSc, Eq.~E4).  
The fitting function of Eq.~(17) tracks the numerical
calibration to better than the $1\%$ level.
%CAUTION FIXED NUMBER
}
\label{fig:kdfit}
\end{figure}

{}From the tight-coupling expansion,
we expect the damping tail to scale as
$\exp[-(\ell/\ell_D)^2]$.
Including the effect of a finite last scattering surface
and the conversion from $k$
to $\ell$ makes the damping function less steep.  We find that through the
first two decades of damping in power, the function ${\cal D}_\ell$ calculated
in the last section can be approximated as
\begin{equation}
{\cal D}_\ell = \exp[ -(\ell/\ell_D)^{m} ].
\label{eqn:diffusionform}
\end{equation}
The quantities $ k_D f = \ell_D/r_\theta$ and $m$ are functions of
$\Omega_0 h^2$ and $\Omega_b h^2$ which are power laws at the extreme ends
of parameter space (see Fig.~\ref{fig:kdfit}).  
Recall that $f \approx 0.98$ is obtained by numerical calibration
of the projection relation (see \S \ref{ss-numerical}).
We chose first to fit the $\Omega_bh^2$ dependence.
A simple two-power law fit
\begin{equation}
\begin{array}{rcl}
\ell_D /r_\theta \eal a_1 (\Omega_b h^2)^{0.291} [1 + a_2 (\Omega_b h^2)^{1.80}]^{-1/5}, \\
m \eal a_3 (\Omega_b h^2)^{a_4} [1 + (\Omega_b h^2)^{1.80}]^{1/5},
\end{array}
\label{eqn:diffusionfit}
\end{equation} 
provides a good description of the numerically determined behavior.
The coefficients of this fit are then functions of $\Omega_0h^2$.
We find that they fit single or double power-laws forms,
\begin{equation}
\begin{array}{rcl}
a_1 \eal 0.0396 (\Omega_0 h^2)^{-0.248} [1 + 13.6 (\Omega_0 h^2)^{0.638}], \\
a_2 \eal 1480 (\Omega_0 h^2)^{-0.0606}
	[ 1 + 10.2 (\Omega_0 h^2)^{0.553}]^{-1}, \\
a_3 \eal 1.03 (\Omega_0 h^2)^{0.0335}, \\
a_4 \eal -0.0473 (\Omega_0 h^2)^{-0.0639}.
\end{array}
\label{eqn:diffusionparam}
\end{equation}
Together these fitting functions work to the percent level for 
$0.02 < \Omega_0 h^2 < 0.75$ and $0.005 < \Omega_b h^2 < 0.75$ 
and improve upon the approximate results of \S \ref{ss-analytic}
and \cite{HSc} (Eq.~E4).  

To complete the description of the damping, we need to express explicitly the
conversion of physical to angular space variables through the angular-size
distance $r_\theta$.  The missing ingredient of Eq.~(\ref{eqn:angularsize})
is the comoving distance to the last scattering surface $\eta_0 -\eta_*$.
The horizon scale today can be expressed as an integral over the Hubble
parameter $\eta_0 = \int_0^1 (a^2 H)^{-1}da$.
For $\Omega_\Lambda=0$, this has the exact solutions
\begin{equation}
\eta_0= \cases{ \fract{1}{H_0 (1-\Omega_0)^{1/2}} \ln \left[
	\fract{2 - \Omega_0 + 2(1-\Omega_0)^{1/2}
	(1 + a_{\rm eq}\Omega_0)}{\Omega_0 + 2(1-\Omega_0)^{1/2}
	(a_{\rm eq} \Omega_0)^{1/2}} \right] & $K<0$, \cr
      		\fract{1}{H_0 (\Omega_0-1)^{1/2}} \left[
		\tan^{-1} {\Omega_0^{1/2} (\Omega_0 - 1)^{-1/2}
			\over 2 a_{\rm eq}^{1/2}
			   (1+a_{\rm eq})^{1/2}}
	    -   \tan^{-1} {2 + 2 a_{\rm eq} - \Omega_0 - 2 a_{\rm eq}
			   \Omega_0 \over 2 (1 + a_{\rm eq}) 
				    (\Omega_0 -1 )^{1/2}} \right]
					     & $K>0$, \cr}
\label{eqn:horizonk}
\end{equation}
while for $K=0$, the following form,
\begin{equation}
\eta_0 = 2 (\Omega_0 H_0^2)^{-1/2}
	 [ (1 + a_{\rm eq})^{1/2} + a_{\rm eq}^{1/2} ]
	 (1 - 0.0841 \ln\Omega_0),
\label{eqn:horizonl}
\end{equation}
fits the integral over the region $0.1 \simlt \Omega_0 \le 1$ and
$0.3\simlt h$ to better than 1\% accuracy. 
Here
\begin{equation}
a_{\rm eq} = 4.17\times 10^{-5}(\Omega_0 h^2)^{-1}
	(T_\gamma/2.728{\rm K})^4
\end{equation}
is the scale factor at matter-radiation equality.
Finally, the horizon at last scattering where $\tau(\eta_*)=1$ takes the form
\begin{equation}
\eta_* = 2(\Omega_0 H_0^2)^{-1/2} [(a_* + a_{\rm eq})^{1/2} - a_{\rm eq}],
\label{eqn:horizonls}
\end{equation}
where (\cite{HSc}, Eq.~E1)
\begin{equation}
\begin{array}{rcl}
z_* \equiv a_*^{-1}-1 \eal 1048 [1 + 0.00124 (\Omega_b h^2)^{-0.738}]
        [1 + b_1 (\Omega_0 h^2)^{b_2} ] ,\\
b_1 \eal 0.0783 (\Omega_b h^2)^{-0.238} [1+39.5(\Omega_b h^2)^{0.763}]^{-1}
,\\
b_2 \eal 0.560 [1+21.1(\Omega_b h^2)^{1.81}]^{-1} ,
\end{array}
\label{eqn:recombfit}
\end{equation}
is a fit to the redshift of recombination.

\subsubsection{Reionization Damping Envelope}

Reionization damping depends on two parameters, the total optical depth
$\tau$ and the angular scale subtended
by the horizon at last scattering during the reionization epoch
$\theta_r \sim \ell_r^{-1}$.  The asymptotic values given in 
Eq.~(\ref{eqn:reionenva})
are highly accurate and thus we need only search for an interpolating
function around $\ell_r$.  The following form fits the behavior 
in ${\cal R}_\ell^2$ to better than 1\% for late reionization 
\begin{equation}
{\cal R}_\ell^2 = {1 - e^{-2\tau} \over 1 + c_1 x
			   + c_2 x^2
			   + c_3 x^3
			   + c_4 x^4} + e^{-2\tau},
\label{eqn:reionenvf}
\end{equation}
with $x = \ell/(\ell_r+1)$ and $ c_1 =  -0.276$, 
     $ c_2 = 0.581$, 
     $ c_3 = -0.172$,
     $ c_4 = 0.0312$.
Even the more extreme case of early reionization to a low ionization level is
described well at the couple of percent level (see Fig.~\ref{fig:reion}).
High precision in the large optical depth limit is unnecessary since secondary
anisotropies dominate in this limit.

The parameter $\ell_r = r_\theta(\eta_r)/\eta_r$ involves the 
visibility-weighted horizon at reionization 
\begin{equation}
\eta_r = {\int d\eta\ \eta\,(\dot \tau e^{-\tau}) \over
	  \int d\eta\ (\dot \tau e^{-\tau})},
\label{eqn:horizonr}
\end{equation}
where the optical depth functions can be obtained by noting that
\begin{equation}
\dot\tau \equiv n_e \sigma_T a = 2.304 \times 10^{-5}{\rm Mpc^{-1}}
	(1 - Y_p) \Omega_b h^2 (1+z)^2 x_h.
\label{eqn:taudot}
\end{equation}
Here $Y_p \approx 0.23 $ is the primordial helium mass fraction 
and recall that $x_h$ is the hydrogen ionization fraction and
we assume that helium is not ionized.
It is useful to note that for low redshifts $z_r \ll 100$ and
constant ionization fraction, the optical depth may be integrated
analytically to give
\begin{equation}
\tau = 4.61 \times 10^{-2} (1-Y_p) x_h {\Omega_b h \over \Omega_0^2}
  \left\{2-3\Omega_0+(1+\Omega_0 z_r)^{1/2}(\Omega_0z_r+3\Omega_0-2) \right\}
\end{equation}
when $\Lambda=0$ and
\begin{equation}
\tau = 4.61 \times 10^{-2} (1-Y_p) x_h {\Omega_b h \over \Omega_0}
  \left\{ [1- \Omega_0 + \Omega_0(1+z_r)^3]^{1/2} - 1 \right\}
\label{eqn:opticaldepth}
\end{equation}
when $K = 0$.
For higher redshifts, the contribution of the radiation to the expansion
rate can make a few percent or greater correction.

\section{Cosmological Information}
\label{sec-information}

Armed with the calibration of the effects of diffusion
and reionization damping, we can now examine the information, both
on cosmological parameters and models for structure formation, embedded
in the small-scale anisotropy spectrum.  We begin with a discussion
of the assumptions that render diffusion and reionization damping
model-independent for most models of structure formation (\S 
\ref{ss-assumptions}).  By
removing the effects of damping in such models, 
one uncovers striking signatures
that contain essential information on the nature of fluctuations 
in the early universe (see also \cite{HWLong}).  
For illustrative purposes, we often employ variants of the standard
CDM model, scale invariant 
initial adiabatic fluctuations $k^3 |\Psi(0,k)|^2 = $const. in
an $\Omega_0=1$, $h=0.5$, $\Omega_bh^2=0.0125$ universe.

Baryon drag, which enhances alternate acoustic peaks (\S\ref{ss-baryon}),
can help separate adiabatic from isocurvature fluctuations, an important
step in distinguishing inflationary models from cosmological defect models
(see \cite{HWLong,HSW}).  
It also probes the gravitational potential at last scattering.
The underlying amplitude of the oscillations extracts information on the
evolution of the gravitational potentials at horizon crossing through the
potential envelope (\S\ref{ss-potential}).

The diffusion damping and reionization damping envelopes are themselves
interesting because they provide essentially model-independent information
about cosmological parameters, mainly the curvature of the universe
(\S \ref{ss-curvature})
and the epoch and extent of reionization (\S\ref{ss-reionization}).  
In this section, we systematically treat these applications of the results
{}from the damping calibration in \S\ref{sec-calculation}.

\subsection{Model Assumptions}
\label{ss-assumptions}

We begin by examining the conditions under which the diffusion damping and
reionization damping envelopes are model-independent to expose general
guidelines for their use.

Only acoustic {\it oscillations} are damped by diffusion.  This leaves
untouched e.g.~offsets in the zero point of the oscillations or anisotropies
generated between the last scattering surface and the observer.
In the former case, the $-\Psi$ offset provided by the potential is not damped
because it represents gravitational redshifts which are picked up by the
photons even as they diffuse in and out of potential wells. 
The baryons provide an inertia to the photon-baryon fluid which further
offsets the oscillation.  The Compton drag of the baryons on the photons
increases the photon temperature inside gravitational potential wells by
$-R\Psi$ leading to a zero-point shift that is not damped by diffusion for
similar reasons.
Together the redshift and drag effects explain why in the estimates of
\S\ref{ss-analytic}, it is $\Theta_0+(1+R)\Psi$ which suffers damping
and not $\Theta_0$. 

The time evolution of the potentials causes a shift of order $\ddot \Psi/k^2$ 
[see \cite{HWLong}, Eq.~(25)].
If $R\gg |\ddot \Psi/k^2\Psi|$, it is negligible in comparison to baryon drag.
Generally $\Psi$ varies on the order of an expansion time such that
$|\ddot \Psi/k^2\Psi| = {\cal O}[(k\eta)^{-2}] \ll 1$ 
for scales well inside the horizon at last scattering: $k\eta\gg 1$.
Mixed terms of the order $R\dot \Psi/k$ also exist but are again generally
smaller than the $R\Psi$ term.
Since the intrinsic acoustic amplitude is of order the gravitational potential 
at sound horizon crossing $\Psi(k,r_s^{-1})$, the diffusion damping signature
dominates over the undamped term if
\begin{equation}
\left| \fract{\Psi(k_D,r_s^{-1})}{\Psi(k_D,\eta_*)} \right|
	\simgt R 
\end{equation}
and
\begin{equation}
k^2 \left| \fract{\Psi(k_D,r_s^{-1})}{\ddot \Psi(k_D,\eta_*)} \right|
\simgt 1 ,
\end{equation}
which are generally satisfied by models whose potentials do not grow
significantly well within the sound horizon.  Notice that no assumption
of coherence in the oscillation is necessary (\cite{Magetal}).

In principle, there is also a model-dependent effect since in the discussion
above we have implicitly assumed a two-step process: first the acoustic
oscillations are formed and then they are damped.
This is generally called in the literature a ``passive'' approximation
(\cite{Albetal}).
If the model possesses a strongly time-varying potential inside the horizon,
the underlying acoustic oscillations could still be forming as the diffusion
length overtakes the wavelength.  Usually this is a small effect, since
most of the damping occurs at the instant of recombination so that the
fluctuations generated during this short time are small. 

Finally anisotropies generated between recombination and the present could be
larger than the intrinsic acoustic signal, especially in the damping tail.
This could occur due to the linear (\cite{Kai84}) and non-linear
(\cite{SZ,Vis}) Doppler effects in a reionized universe or time variations in
the potential along the line of sight (\cite{SacWol,RS,KaiSte}).  
In models such as cold dark matter (CDM) with a near 
scale-invariant spectrum
of adiabatic initial fluctuations, this is not a worry.  The lack of excessive
small scale power in the model makes early reionization and/or small-scale
non-linearities that are responsible for such effects unlikely 
(\cite{CebBar,SelRS,HuWhi}).

These types of considerations also apply to the reionization damping function
${\cal R}_\ell$ calculated in \S \ref{ss-numerical}.
By construction, this function isolates the rescattering damping effect
during reionization and ignores any secondary effects that may
regenerate fluctuations.  Again, the Doppler effect due to
the relative velocity of the electrons with respect to the CMB can
regenerate fluctuations significantly if {\it both} the peculiar
velocities and the optical depth are large.  We examine this
effect more closely in \S\ref{ss-reionization}.

In summary, the damping function ${\cal D}_\ell$ accurately describes
the model-independent damping of acoustic
{\it oscillations} and the reionization damping function ${\cal R}_\ell$
does the same for the rescattering damping of {\it primary} anisotropies.  
In models such as CDM with no excess small scale power and hence
relatively late reionization and small secondary effects, 
their behavior will be clearly manifest in the observable spectrum.  
In models where this is not true, it merely describes the behavior 
of a component of the total anisotropy and other effects must be 
taken into account to extract the information embedded in the 
observed anisotropy.

\begin{figure}[t]
\begin{center}
\leavevmode
\epsfxsize=3.5in \epsfbox{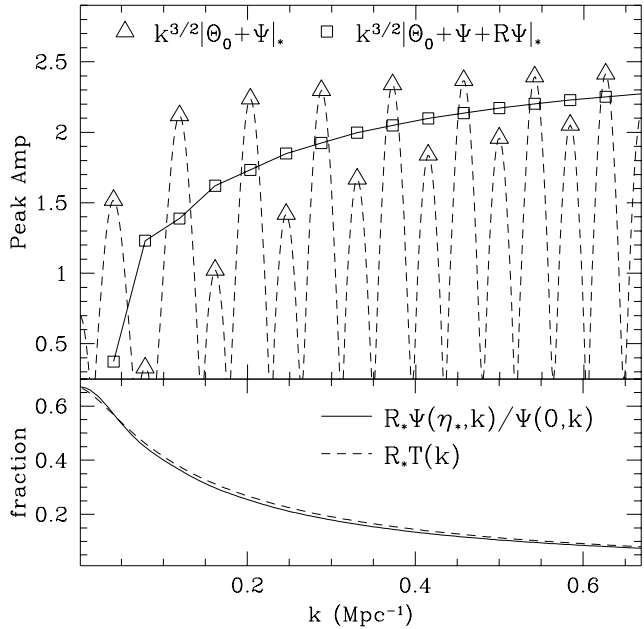}
\end{center}
\caption{Baryon drag and its potential dependence.  Baryon inertia
in the fluid displaces the zero point of the temperature oscillations
leading to alternating peak heights as a function of scale at last 
scattering.  The magnitude of the displacement is $R_*\Psi(\eta_*)$, and by
removing it the monotonic variation of heights due to the
potential envelope is uncovered (upper panel). The fractional
effect is of order $R_*\Psi(\eta_*,k)/\Psi(0,k)$ and
can be adequately described by the matter transfer function $T(k)$
(lower panel).
The model here is 
CDM with $\Omega_0=1$, $h=1$ and $\Omega_bh^2 = 0.025$.}
\label{fig:drag}
\end{figure}

\subsection{Uncovering the Baryon Signature}
\label{ss-baryon}

Baryons create a distinct acoustic signature due to the drag effect discussed
in \S\ref{ss-assumptions}.  By providing inertia to the fluid, they enhance 
compressions over rarefactions inside potential wells.
Aside from providing a means to measure the baryon content, it also
distinguishes between between the two phases through the difference in peak
amplitudes between successive peaks.
In turn this distinction provides one of the most striking and robust ways to
distinguish adiabatic inflationary fluctuations from their isocurvature
counterparts, generated perhaps by cosmological defects (\cite{HWLong,HSW}).
Unfortunately damping and projection effects serve to obscure this signal.
By deconvolving these effects with the results and methods of
\S\ref{ss-numerical}-\ref{ss-fitting}, one can uncover this important
signature.

Let us first examine the intrinsic effect.  In Fig.~\ref{fig:drag}, 
we show an example from a solution of the tight-coupling oscillator 
equation Eq.~(\ref{eqn:oscillator})
under the metric fluctuations of an $\Omega_0 = 1$, $h=1$, $\Omega_b h^2
=0.025$ CDM model. 
Displayed is the effective temperature fluctuation of the peaks (triangles),
connected by the full function to guide the eye.  
To demonstrate that the alternating height effect is due to 
baryon drag, 
we add $R_*\Psi(\eta_*,k)$ to each peak (squares), where $R_*=R(\eta_*)$.
Notice that this eliminates the alternation, leaving
the peak heights to smoothly vary 
in a manner described by the ``potential
envelope'' discussed in the next section.  
Since the intrinsic amplitude of the oscillations is of order the
potential before sound horizon crossing, the fractional effect 
is of order $R_*\Psi(\eta_*,k)/\Psi(0,k)$. 

Since the effect depends
on the potential at last scattering $\Psi(\eta_*,k)$, it also provides 
a probe of the matter fluctuations at that epoch. 
Under the CDM scenario, the potential does not evolve significantly
between recombination and the present so that the baryon drag
effect also reflects the matter fluctuations today.  The fractional
effect becomes 
$R_*\Psi(\eta_*,k)/\Psi(0,k) 
\approx R_*\Psi(\eta_0,k)/\Psi(0,k) = R_* T(k)$, 
where $T(k)$ is the matter transfer function (\cite{BBKS}),
\begin{equation}
T(q) = {\ln(1 + 2.34q) \over 2.34 q}
	[ 1 + 3.89 q + (16.1q)^2 + (5.46q)^3 + (6.71 q)^4 ]^{-1/4},
\label{eqn:BBKS}
\end{equation}
with $q = (k/{\rm Mpc}^{-1}) (T_\gamma/2.7{\rm K})^2 (\Omega_0 h^2)^{-1}$.
In Fig.~\ref{fig:drag} (lower panel), we show that $RT(k)$ accurately
tracks the effect and provides a potential 
consistency check with large scale
structure today.  As we shall see in the next section, the 
fall of the fractional baryon
drag effect and the rise of the potential envelope
are intimately related through the matter-radiation equality epoch.

\begin{figure}[t]
\begin{center}
\leavevmode
\epsfxsize=3.5in \epsfbox{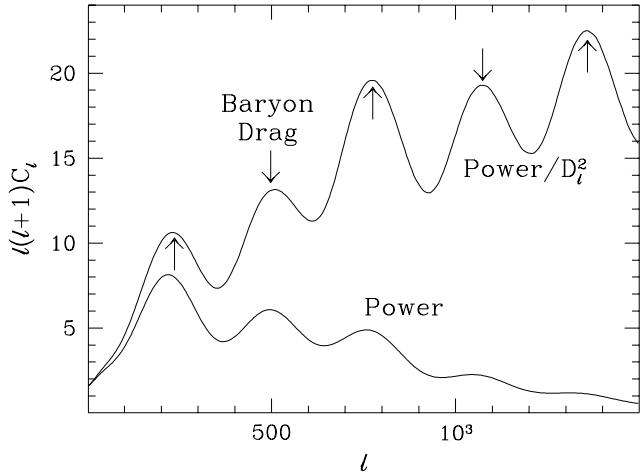}
\end{center}
\caption{Uncovering Baryon Drag in a low baryon universe.
Diffusion damping obscures the baryon drag signal especially in a
low baryon universe (here $\Omega_b h^2=0.075$ in an otherwise
standard CDM model).  Employing the numerical calibration of the 
damping tail, we recover the alternations.}
\label{fig:undampdrag}
\end{figure}

The magnitude of the baryon drag effect in the observable anisotropy spectrum
is reduced by inclusion of the dipole term and smoothing by projection, but
mainly by diffusion damping.  If the baryon content is low, the intrinsic
magnitude of the effect is small and diffusion damping may cause the peak
heights to monotonically decrease rather than alternate 
(see Fig.~\ref{fig:undampdrag}).
Given the calibration of the diffusion damping behavior in
\S\ref{ss-numerical}, we can invert this filter.
In Fig.~\ref{fig:undampdrag}, we show that multiplying the spectrum by
${\cal D}_\ell^{-2}$ uncovers the alternating peaks even for $\Omega_b h^2$
significantly lower than the standard big bang nucleosynthesis prediction.  
In practice, removing the damping behavior will require knowledge of
$\Omega_0 h^2$ and $\Omega_b h^2$ either from external measurements or
consistency checks (see \cite{HWLong}) as well as measurement of the curvature
from the CMB.

\begin{figure}[t]
\begin{center}
\leavevmode
\epsfxsize=3.5in \epsfbox{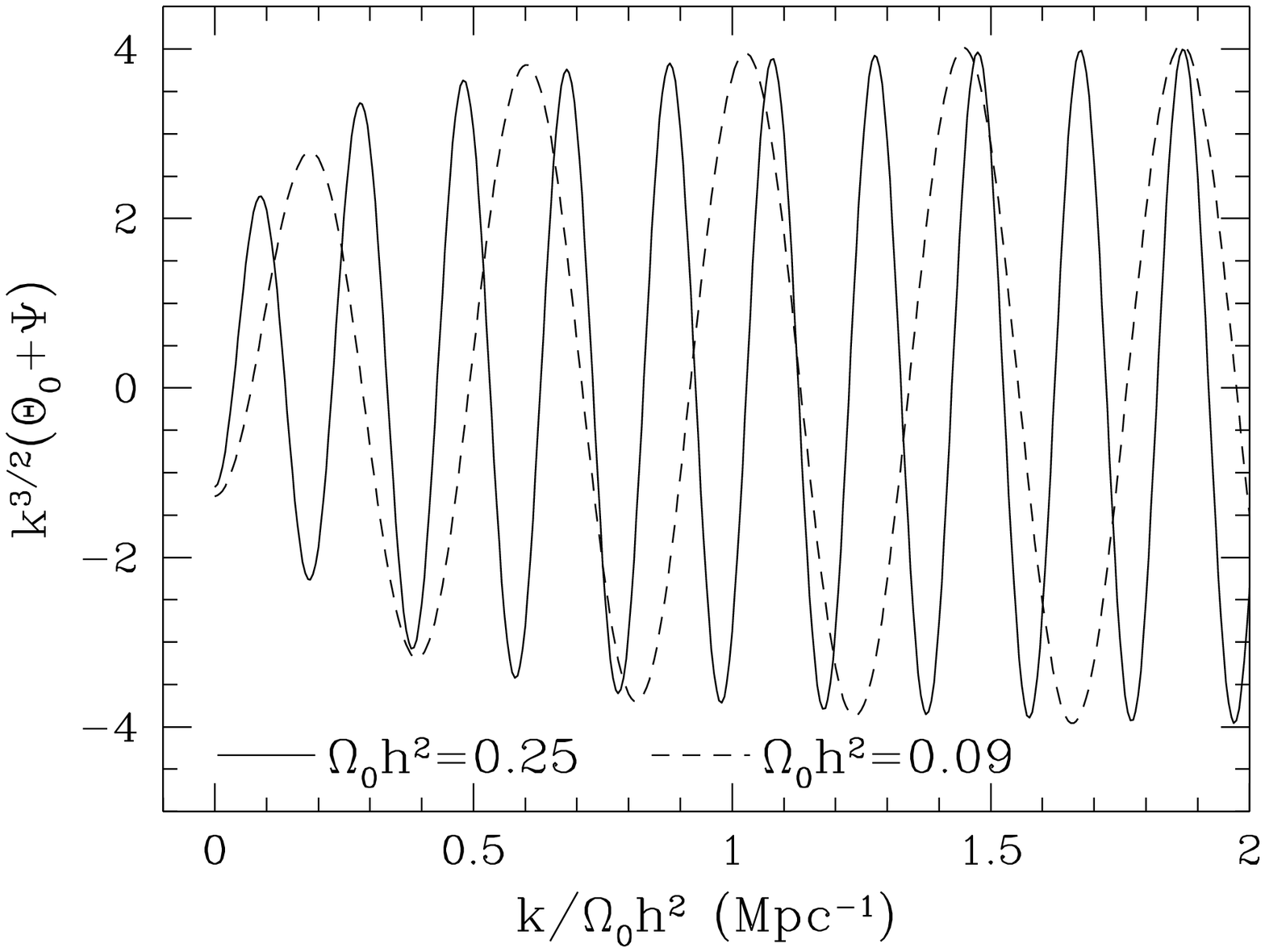}
\end{center}
\caption{Potential Envelope.  Decay of the potential due to the self
gravity of the photon-baryon fluid drives the oscillator. Comparing
two CDM models with differing matter to radiation ratios $\Omega_0 h^2$,
we see that the oscillations are multiplied by an envelope that 
depends on the equality scale $k_{\rm eq} \propto \Omega_0 h^2$.}
\label{fig:envelope}
\end{figure}

\subsection{Determining the Potential Envelope}
\label{ss-potential}

Gravitational potential perturbations drive acoustic oscillations affecting
their amplitude and phase.  The effect on the phase can be used to uncover
information about the origin of fluctuations in an inflationary epoch or phase
transition (\cite{HSb,CriTur,HWLong}).
Here we treat their effects on the {\it amplitude\/} of the intrinsic 
oscillations, unobscured by the presence of diffusion damping.
This can be obtained from an observed spectrum by the techniques of
\S\ref{sec-calculation} and is also useful for constraining the curvature
(see \S\ref{ss-curvature})

As an example of the driving process, let us consider the case of adiabatic
fluctuations.  The self-gravity of the photon-baryon fluid drives its own
oscillations through a feedback mechanism at sound horizon crossing.  
Photon pressure prevents gravitational collapse inside the sound horizon
leading to a decay in the self-generated gravitational potential.
This decay is timed such that it leaves the oscillator in a highly compressed
state leading to correspondingly large amplitude acoustic oscillations
(see \cite{HWLong} for further description).
If the self-gravity of the photons and baryons dominate, the amplitude of the
oscillation is enhanced from gravitational redshifts by $-2\Psi$ which combined
with the Sachs-Wolfe effect (\cite{SacWol}) of ${1\over 3}\Psi$ yields a net
result of $-{5 \over 3}\Psi$, i.e.~the amplitude of the oscillation should be
5 times the large-angle Sachs-Wolfe plateau.
Inclusion of neutrinos and the matter-radiation transition modify this result
to $5 (1 + {4 \over 15}f_\nu)^{-1}$, where the neutrino density fraction is
$f_\nu = \rho_\nu/(\rho_\nu+\rho_\gamma)$ (\cite{HSc}, Eq.~B9).
This driving effect only operates if the self-gravity of the photon-baryon
fluid dominates at sound horizon crossing.
Large scales cross the sound horizon in the matter-dominated epoch and do not
suffer this effect.
Thus the scale that crosses the horizon at matter-radiation equality
$k_{\rm eq}$ marks the transition between the two asymptotic regimes.

The critical scale
$k_{\rm eq}=(2\Omega_0 H_0^2/a_{\rm eq})^{1/2}\propto \Omega_0 h^2$ 
provides the CMB with sensitivity to the parameter $\Omega_0 h^2$.
This is similar to the more familiar effect of equality on the matter power
spectrum [see Eq.~(\ref{eqn:BBKS})] but note that fluctuations increase rather
than decrease upon crossing $k_{\rm eq}$.
Fig.~\ref{fig:envelope} shows that the potential envelope that governs the
amplitude is indeed a function of $k/k_{\rm eq}\propto k/\Omega_0 h^2$.
Potentially, this effect can also probe the neutrino mass through its effect
on $k_{\rm eq}$ (see \cite{SelBer,MaBer,DodGatSte}).

The remaining subtlety is that the presence of baryons makes acoustic
oscillations decay adiabatically.  Notice that the tight-coupling equation
(\ref{eqn:oscillator}) describes an acoustic oscillator with effective mass
of $(1+R)$.
The adiabatic invariant for such an oscillator is the energy/frequency.
This requires that temperature fluctuations decay as $(1+R)^{-1/4}$ and
dipole or Doppler contributions to decay as $(1+R)^{-3/4}$.  
The amplitude of the potential envelope thus gains a baryon dependence set by
the value of $R$ at recombination (\cite{HSc}).  

\begin{figure}[t]
\begin{center}
\leavevmode
\epsfxsize=3.5in \epsfbox{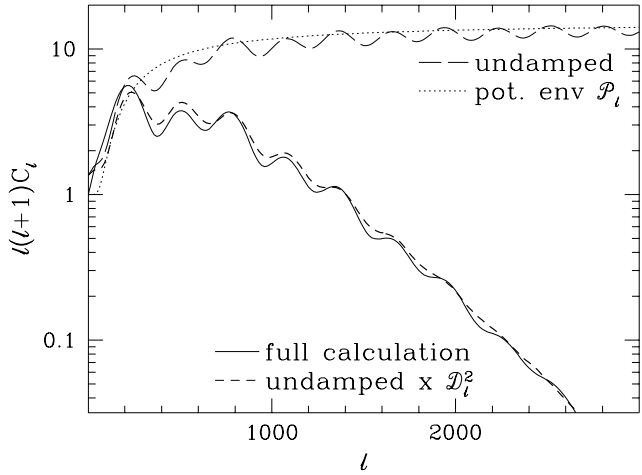}
\end{center}
\caption{Uncovering the Potential Envelope.  The potential envelope 
is obscured by diffusion damping.  By numerically removing the
damping, one sees that the intrinsic fluctuations follow the
analytic estimates of ${\cal P}_\ell$ reasonably well. By multiplying
by the numerically-calibrated damping function ${\cal D}_\ell^2$, one
recovers the form of the full calculation even at very small angles.
The model here is standard CDM.
}
\label{fig:scdm}
\end{figure}

The full potential envelope in power can be roughly described by
\begin{equation}
{\cal P}_\ell = 1 + A \exp(-1.4 \ell_{\rm eq}/\ell),
\label{eqn:potenv}
\end{equation}
for a scale-invariant spectrum.
Here $\ell_{\rm eq} = k_{\rm eq} r_\theta$ and the amplitude $A$ is fixed
by the asymptotic expression
\begin{equation}
A = {25} (1+{4\over 15}f_\nu)^{-2}
    \ {(1+R_*)^{-1/2} + (1+R_*)^{-3/2}\over 2} - 1,
\label{eqn:potenvamp}
\end{equation}
where we have combined the temperature and Doppler effects in quadrature.
Tilting the primordial spectrum produces an analogous tilt in 
${\cal P}_\ell$.  The integrated Sachs-Wolfe effect 
(\cite{SacWol}) in open and $\Lambda$ models, also gives
${\cal P}_\ell$ large-angle contributions (see also \cite{HSS}).

We show an example in Fig.~\ref{fig:scdm}.  The upper curves show a
calculation with the effects of diffusion damping removed compared with
the potential envelope of Eq.~(\ref{eqn:potenvamp}).  Note that the form
of the envelope roughly traces power in the fluctuations.
The bottom curves show how diffusion damping obscures the signature and
tests the damping calibration of \S\ref{ss-numerical} in a realistic context.
By multiplying the undamped calculation by ${\cal D}_\ell^2$ one regains,
to reasonable accuracy, the result of a full CDM calculation encorporating
diffusion damping.

Thus the obscuring effects of diffusion damping can be removed to
extract the potential envelope of acoustic oscillations.  This provides
information on the evolution of the metric fluctuations
as they cross the sound horizon which may help unravel information
about the nature of such fluctuations in the general case and
the scale of matter-radiation equality in the adiabatic case.

\begin{figure}[t]
\begin{center}
\leavevmode
\epsfxsize=3.5in \epsfbox{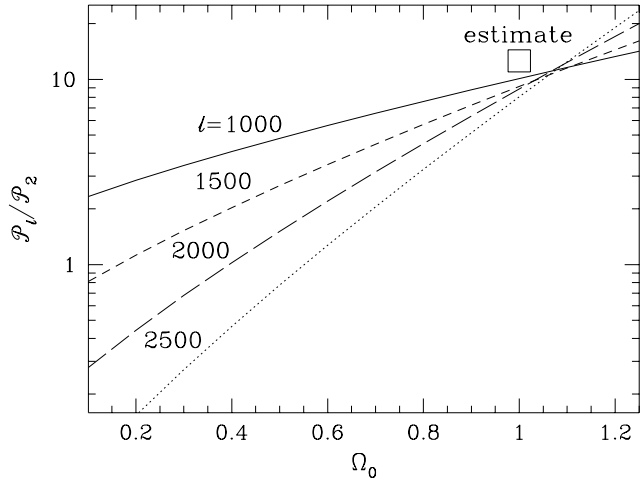}
\end{center}
\caption{Constraining $\Omega_0$ with the damping tail. 
By measuring
the anisotropy power in at some scale $\ell$ in the damping tail 
(here averaged over 10\% in $\ell$) and comparing it to a reference
scale (here $\ell=2$), one determines the ratio of intrinsic
powers ${\cal P}_\ell/{\cal P}_2$
before damping necessary to reproduce the observation 
(here $\Omega_0=1$ in standard CDM).
Since this is a strong function of the assumed $\Omega_0$, only order
of magnitude knowledge of the model-dependent intrinsic power 
is needed (e.g. square, estimated from Eq.~(32))
%CAUTION FIXED NUMBER
to reject values of $\Omega_0$. Multiple measurements
in the damping tail largely removes this ambiguity (curve intersection).
For simplicity, we have fixed $h=0.5$, $\Omega_bh^2=0.0125$ and
$\Omega_\Lambda=0$.}
\label{fig:omega}
\end{figure}

\subsection{Constraining the Curvature}
\label{ss-curvature}

The angular scale of diffusion damping $\ell_D$ 
provides a clear feature by 
which a classical angular-size distance test of the curvature can
be made by comparison with the corresponding physical scale $k_D$
(\cite{HuWhi}).
In models with simple acoustic peak features which can also be used
for this test, this provides a consistency check on the curvature, 
important if the baryon content or thermal history of the universe
is unknown or anomalous (\cite{HWLong}).  In models where the peak 
signature is more complicated
or non-existent (\cite{Albetal}), it may serve as the primary means 
of measuring the curvature.  

In principle, the curvature is constrained by the simple absence or
presence of small scale power.  In an open universe, geodesic deviation
moves the diffusion tail in angular space to smaller angles leading
to more power on small scales.  In practice, its application is complicated
by secondary effects in the foreground and lack of {\it a priori} knowledge 
about the intrinsic amplitude of fluctuations before damping.
The former is unlikely to be an obstacle in models with no strong
non-linearities at small scales, in which the acoustic signal from
recombination is the dominant contribution to the anisotropy.  

Lack of knowledge of the intrinsic amplitude of oscillations limits the
precision by which the curvature can be measured from the damping tail. 
The intrinsic amplitude is given by the potential envelope ${\cal P}_\ell$
discussed in the last section.  
Given that diffusion damping is exponential in $\ell$, it takes only
a rough estimate of ${\cal P}_\ell$ to yield interesting constraints
on the curvature.  Furthermore if ${\cal P}_\ell$ is a slowly varying
function compared with ${\cal D}_\ell$, measurement of the power
on several scales in the damping tail can remove the ambiguity.

We show an example in Fig.~\ref{fig:omega}.  Here we assume that
the underlying spectrum is of standard CDM which sets 
$\Omega_0=1$.  By comparing the power at some
scale $\ell$ in the damping tail to some reference scale, here $\ell=2$,
\begin{equation}
{\ell(\ell+1)C_\ell \over 6C_2} = {{\cal D}_\ell^2 \over {\cal D}_2^2}
	{{\cal P}_{\ell} \over {\cal P}_2},
\end{equation}
and by using the fitting formula for ${\cal D}_\ell$ of
Eq.~(\ref{eqn:diffusionfit}), one can determine the intrinsic ratio of power
${\cal P}_{\ell}/{\cal P}_2$ as a function of $\Omega_0$ needed to
reproduce the measurement.
We have ignored here the suppression of power from ${\cal R}_\ell$ as it is
generally negligible for our purposes here.
Because the damping multipole $\ell_D$ is a strong function of $\Omega_0$,
the amount of intrinsic power required increases steeply with $\Omega_0$.
Thus even the crude estimate of the CDM potential envelope of
Eq.~(\ref{eqn:potenv}) is more than sufficient to distinguish between
interesting values of $\Omega_0$ (see Fig.~\ref{fig:omega}, square).  
As the slope of the curves reflect, the further into the damping tail one
can measure, the more powerful the test becomes.
Of course for $\ell\gg\ell_D$, the signal also drops exponentially and hence
is difficult both to measure and separate from secondary effects.

By measuring more than one scale in the damping tail, one obtains a
consistency check on the curvature constraint.  If the $\ell$ dependence
of ${\cal P}_\ell$ is weak, as is the case for CDM-like scenarios
(see Fig.~\ref{fig:scdm}), then the predictions for the intrinsic power
must intersect near the actual value of $\Omega_0$.
This implements the damping-tail shape test proposed in \cite{HWLong}
to remove the model dependence of the curvature constraint. 
Even if only upper limits exist from CMB measurements at small scales,
lower limits on $\Omega_0$ can be obtained with reasonable assumptions
on the amount of intrinsic power in small scale fluctuations.

\begin{figure}[t]
\begin{center}
\leavevmode
\epsfxsize=3.5in \epsfbox{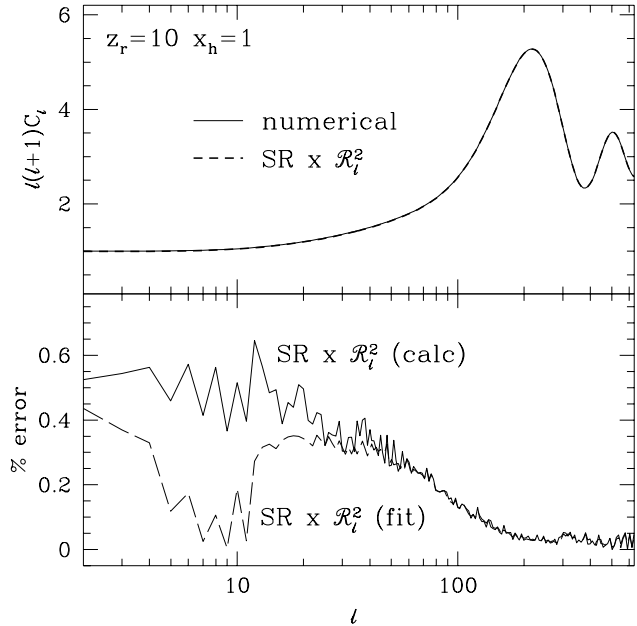}
\end{center}
\caption{Reionization damping in standard CDM.  Damping described by the
envelope ${\cal R}_\ell$ is the main effect of late reionization in CDM
type models.  Hence employing either the numerical calibration of
${\cal R}_\ell$ and the fit to it from Eq.~(24) to filter the results of a
%CAUTION FIXED NUMBER
standard recombination (SR, no reionization) calculation approximate the
full calculation to better than 1\% in power.  The scatter at low $\ell$
is a numerical artifact from finite sampling of the $C_\ell$ integral in
$k$-space (see Eq.~(4)).
% CAUTION FIXED NUMBER
}
\label{fig:z10}
\end{figure}

\subsection{Examining Reionization}
\label{ss-reionization}

Even late reionization produces potentially observable consequences for
precise measurements of the CMB.  In a standard CDM model, the optical
depth ranges from 1-3\% between $5 < z_r < 10$ leading to a 2-6\% effect in
the anisotropy power spectrum.  
For these low optical depths, it is likely that the main effect of
reionization is the rescattering damping calculated in \S\ref{ss-numerical}.
In this case, two cosmological quantities are potentially extractable from the
spectrum, the total optical depth and the horizon size at last scattering
during the reionized epoch.  In practice, extracting accurate results will be
hampered by cosmic variance at large angles and the close degeneracy between
changes in the spectrum due to the normalization and late reionization at small
angles.

In Fig.~\ref{fig:z10}, we show how well the numerical calibration of
\S\ref{ss-numerical} and the fitting formula of \S\ref{ss-fitting}
Eq.~(\ref{eqn:reionenvf}) reproduce the full effect of late reionization.
The accuracy achieved is always better than a percent in power and increases
toward small scales where the reionization signal is the largest.
With the high precision achievable by the next generation satellite
experiments, it is conceivable that the CMB spectrum can probe even such
relatively late reionization.

\begin{figure}[t]
\begin{center}
\leavevmode
\epsfxsize=3.5in \epsfbox{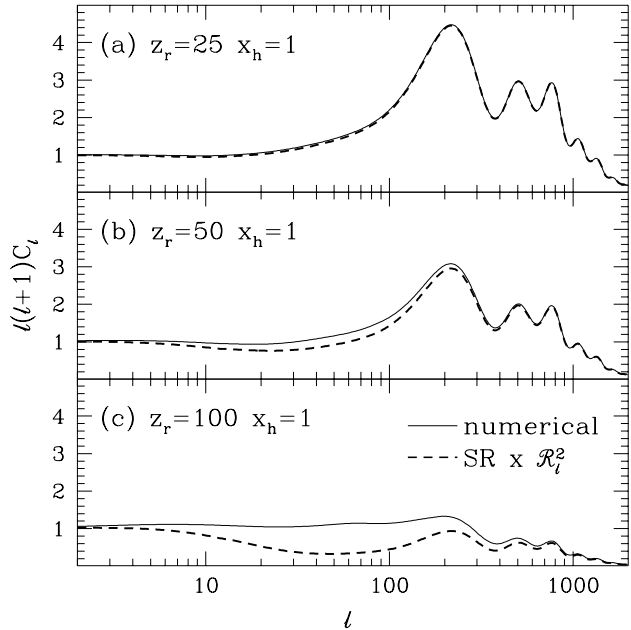}
\end{center}
\caption{Reionization and the Doppler Effect.  For early ionization,
the Doppler effect due to the relative electron-photon velocity can
regenerate fluctuations around the horizon scale at the last scattering
epoch.  By comparing the standard recombination (SR) result filtered
by reionization damping ${\cal R}_\ell^2$ to the full calculation,
we can uncover such effects.}
\label{fig:cdmreion}
\end{figure}

If reionization occurs earlier, such that the optical depth is higher and/or
non-linear effects dominate, then its effect on the CMB can be even more
significant.  Fluctuations are not only erased but also regenerated.
As an example, consider the Doppler effect from the relative velocity of
the electrons with respect to the CMB generated as the baryons fall into
dark matter potential wells.
Its effect peaks near the horizon at last scattering due to competing effects.
Velocity flows are only generated inside the horizon.  Yet on small scales, 
photons last scatter against many crests and troughs of the velocity
perturbation leading to a strong cancellation damping of the Doppler effect
(\cite{Kai84}).  
By employing the rescattering damping function ${\cal R}_\ell$ from
\S\ref{ss-numerical}, we isolate this effect in Fig.~\ref{fig:cdmreion}.
For the higher optical depth cases, the Doppler effect is clearly apparent
as an excess of fluctuations from that predicted by ${\cal R}_\ell$.
On scales much smaller than the horizon at the last scattering epoch, simple
analytic approximations exist for this effect (\cite{Kai84,HuWhi}).
In a CDM model where the optical depth is likely to be $\tau\simlt 1$, such
small scale effects are masked by larger primary anisotropies until well into
the damping tail.  

More complicated rescattering damping can occur if the reionization is patchy.
Although one cannot directly apply the results of our damping calibration to
this case, basic elements uncovered such as the dependence of damping on the
horizon scale can be applied to this case as well.
Non-linear effects can also create fluctuations through the Doppler effect but
these are generally small in a model like CDM without excessive small scale
power (but see \cite{Aghetal}).

\section{Conclusions}

Prospects for measuring the small scale CMB anisotropy spectrum are bright,
especially in light of the approval of two new satellite missions, 
MAP from
NASA and Cobras/Samba from ESA, and the funding of ground based interferometers.
If foregrounds, systematic and secondary effects are small or can be removed,
and the inflationary CDM model is correct, much cosmological information can
be extracted from the damping tail of CMB anisotropies
(see e.g.~\cite{CobSam}).
Despite the enormous success of this model however, it is quite possible that
what is found there will come as a surprise to the current orthodoxy in
cosmological modeling.  
In preparation for this possibility we have here, and in \cite{HWLong},
attempted to construct the spectrum out 
of fundamental physical effects that are likely to be the elements in any
future model which successfully explains the observations.

The basic elements uncovered here represent a series of 
numerically calibrated transfer functions
that describe the linear processing of fluctuations: the diffusion damping
envelope, the reionization damping envelope, the potential envelope
and the baryon drag modulation.
The anisotropy spectrum is not merely a snapshot of conditions on
the last scattering surface.  Rather it is a dynamic entity that
bears the mark of fluctuations before
horizon crossing through the acoustic phase (\cite{HWLong}),
at horizon crossing through the potential envelope, at last
scattering through baryon drag, and after last scattering through
the large-angle potential envelope (\cite{SacWol})
as well as the effects of reionization.
Within the present framework of model possibilities, this view of
its structure also creates a system of consistency checks 
by which we can verify 
model assumptions, such as the inflationary or cosmological defect
origin of fluctuations, before proceeding to measure cosmological
parameters and details of the model.  

\noindent{\it Acknowledgments:} W.H. was supported by a grant
from the W.M. Keck Foundation.

\vfill

\noindent{\tt whu@sns.ias.edu}

\noindent{http://www.sns.ias.edu/$\sim$whu}


\clearpage

\begin{thebibliography}{99}

\bibitem[Aghanim et al.~1996]{Aghetal}
 Aghanim, N., Desert, F. X., Puget, J. L. \& Gispert, R. 1996, A\&A, 311, 1
 [astro-ph/9605083]
\bibitem[Albrecht et al.~1995]{Albetal}
 Albrecht, A., Coulson, D., Ferreira, P., \& Magueijo, J. 
 1996, Phys. Rev. Lett., 76, 1413 [astro-ph/9505030]
\bibitem[Bardeen et al.~1987]{BBKS}
 Bardeen, J. M., Bond, J. R., Kaiser, N., \& Szalay, A. S. 1987, \apj, 304, 15
\bibitem[Bersanelli et al.~1996]{CobSam}
 Bersanelli, M. et al. 1996, ESA Phase A Study  \\
 (http://astro.estec.esa.nl/SA-general/Projects/Cobras/report/report.html)
\bibitem[Bond 1996]{Bond}
 Bond, J. R. in Cosmology and Large Scale Structure, ed. by R. Shaeffer
 et al. (Elsevier Science Publishers, Netherlands) in press. 
\bibitem[Bond \& Efstathiou 1984]{BE84}
 Bond, J. R., \& Efstathiou, G. 1984, \apj, 285, L45
\bibitem[Bond \& Efstathiou 1987]{BE87}
 Bond, J. R., \& Efstathiou, G. 1987, \mnras, 226, 665
\bibitem[Bond et al.~1994]{Bonetal}
 Bond, J. R., et al.~1994, Phys. Rev. Lett., 72, 13 [astro-ph/9309041]
\bibitem[Ceballos \& Barcons~1994]{CebBar}
 Ceballos, M. T., \& Barcons, X. 1994, \mnras, 271, 817
\bibitem[Crittenden \& Turok~1995]{CriTur}
 Crittenden, R. G., \& Turok, N. G. 1995, Phys. Rev. Lett., 75, 2642
 [astro-ph/9505120]
\bibitem[Dodelson, Gates \& Stebbins 1995]{DodGatSte}
 Dodelson, S., Gates, E., \& Stebbins, A. 1996, \apj, 467, 10 
 [astro-ph/9509147]
\bibitem[Dodelson \& Turner 1992]{DodTur}
 Dodelson, S., \& Turner, M. S. 1992, Phys. Rev., D46, 3372
\bibitem[Doroshkevich, Zel'dovich \& Sunyaev 1978]{DZS}
 Doroshkevich, A. G., Zel'dovich, Ya. B., \& Sunyaev, R. A. 1978,
 Sov. Astron., 22, 523
\bibitem[Durrer et al.~1996]{Duretal}
 Durrer, R., Gangui, A., \& Sakellariadou, M. 1996, Phys. Rev. Lett., 76, 579
 [astro-ph/9507035]
\bibitem[Efstathiou \& Bond~1987]{EfsBon}
 Efstathiou, G., \& Bond, J. R. 1987, \mnras, 227, 33p
\bibitem[Fixsen et al.~1996]{Fixetal}
 Fixsen, D. J., et al. 1996, \apj, [in press, astro-ph/9605054]
\bibitem[Gunn \& Peterson~1965]{GunPet}
 Gunn, J. E., \& Peterson, B. A. 1965, \apj, 318, L11 
\bibitem[Hu et~al.~1995]{HSSW}
 Hu, W., Scott, D., Sugiyama, N., \& White, M. 1995, Phys. Rev. D., 52, 5498 
\bibitem[Hu, Spergel, \& White~1996]{HSW}
 Hu, W., Spergel, D. N., \& White, M. 1996, [astro-ph/9605193]
\bibitem[HSa]{HSa}
 Hu, W., \& Sugiyama, N. 1995, \apj, 444, 489 (HSa) [astro-ph/9406071]
\bibitem[HSb]{HSb}
 Hu, W., \& Sugiyama, N. 1995, Phys. Rev. D, 51, 2599 (HSb)
 [astro-ph/9411008]
\bibitem[HSc]{HSc} 
 Hu, W., \& Sugiyama, N. 1996, \apj (HSc) [in press, astro-ph/9510117] 
\bibitem[Hu, Sugiyama \& Silk 1996]{HSS} 
 Hu, W., Sugiyama, N., \& Silk, J. 1996, Nature, [in press, astro-ph/9604166] 
\bibitem[Hu \& White~1996a]{HuWhi}
 Hu, W., \& White, M. 1996a, A\&A, [in press, astro-ph/9507060]
\bibitem[Hu \& White~1996b]{HWLong}
 Hu, W., White, M. 1996b, \apj [in press, astro-ph/9602019]
%\bibitem[Jungman et al. 1995]{Jungman}
% Jungman, G., Kamionkowski, M., Kosowski, A. \& Spergel, D. N. 1995, 
 Phys. Rev. Lett., 76, 1007 [astro-ph/9507080]
\bibitem[Jungman et al.~1996]{JunPRD}
 Jungman, G., Kamionkowski, M., Kosowski, A. \& Spergel, D. N. 1996, 
 Phys. Rev. D., 54, 1332 [astro-ph/9512139]
\bibitem[Kaiser~1983]{Kai}
 Kaiser, N. 1983, \mnras, 202, 1169
\bibitem[Kaiser~1984]{Kai84}
 Kaiser, N. 1984, \apj, 282, 374
\bibitem[Kaiser \& Stebbins~1984]{KaiSte}
 Kaiser, N. \& Stebbins, A. 1984, Nature, 310, 391
\bibitem[Kamionkowski, Spergel \& Sugiyama~1994]{KamSS}
 Kamionkowski, M., Spergel, D. N., Sugiyama, N. 1994, \apj, 426, L57
 [astro-ph/9401003]
\bibitem[Ma \& Bertschinger~1995]{MaBer}
 Ma, C. P. \& Bertschinger, E. 1995, \apj, 455, 7 [astro-ph/9506072]
\bibitem[Magueijo et al.~1996]{Magetal}
 Magueijo, J., Albrecht, A., Coulson, D., \& Ferreira, P. 
 1996, Phys. Rev. Lett., 76, 2617  [astro-ph/9511042]
\bibitem[Peebles \& Yu~1970]{PeeYu}
 Peebles, P. J. E. \& Yu, J. T. 1970, \apj, 162, 815
\bibitem[Rees \& Sciama~1968]{RS}
 Rees, M. J. \& Sciama, D. N. 1968, Nature, 217, 511  
\bibitem[Sachs \& Wolfe~1967]{SacWol}
 Sachs, R. K. \& Wolfe, A. M. 1967, \apj, 147, 73
\bibitem[Seljak 1994]{Sel}
 Seljak, U. 1994, \apj, 435, L87 [astro-ph/9406050]
\bibitem[Seljak~1996]{SelRS}
 Seljak, U. 1996, \apj, 460, 549 [astro-ph/9506048]
\bibitem[Seljak \& Bertschinger~1993]{SelBer}
 Seljak, U., Bertschinger, E. 1993, Proceedings of ``Present and Future of
 the Cosmic Microwave Background'', Santander, Spain, ed.~J.L.Sanz,
 E.Martinez-Gonzalez, L.Cayon, (Springer Verlag, Berlin), p.165
\bibitem[Seljak \& Zaldarriaga 1996]{SelZal}
 Seljak, U.  \& Zaldarriaga, M. 1996, astro-ph/9603033
\bibitem[Silk~1968]{Sil}
 Silk, J. 1969, \apj, 151, 459
\bibitem[Sugiyama \& Gouda~1992]{SugGou}
 Sugiyama, N. \& Gouda, N. 1992, Prog. Theor. Phys., 88, 803
\bibitem[Sunyaev \& Zel'dovich~1970]{SZ}
 Sunyaev, R. A. \& Zel'dovich 1970, Ya. B., Astrophys. Space Sci. 7 1
\bibitem[Vishniac 1987]{Vis}
 Vishniac, E. T. 1987, \apj, 322, 597
\bibitem[Weinberg~1972]{Wei}
 Weinberg, S. 1972, Gravitation and Cosmology (Wiley, New York) p. 568
\bibitem[White \& Scott~1996]{WhiSco}
 White, M., \& Scott, D. 1996, \apj, 459, 415
 [astro-ph/9508157]
\bibitem[Wilson 1983]{Wil}
 Wilson, M. L. 1983, \apj, 273, 2

\end{thebibliography}
\end{document}